\documentclass{appolb}
\usepackage{epsfig}
\usepackage{amsmath,amsfonts,amssymb}
\usepackage{t1enc}
\usepackage{cite}

\begin{document}
% \eqsec  % uncomment this line to get equations numbered by (sec.num)
\title{Lepton Number Violation and 
Scalar Searches at the LHC
\thanks{Presented by F. del Aguila at the XXXVII International Conference 
of Theoretical Physics ``Matter To The Deepest: Recent Developments in 
Theory of Fundamental Interactions'', Ustro\'n, Poland, September 1-6, 2013, 
and at ``From Majorana to LHC: workshop on the origin of neutrino mass'', 
Trieste, Italy, October 2-5, 2013, 
and by M. Chala at ``Scalars 2013'', Warsaw, Poland, September 12-16, 2013.}
}
\author{F. del Aguila$^a$\footnote{\tt faguila@ugr.es}, 
M. Chala$^a$\footnote{\tt miki@ugr.es}, 
A. Santamaria$^b$\footnote{\tt Arcadi.Santamaria@uv.es} 
and J. Wudka$^c$\footnote{\tt jose.wudka@ucr.edu} 
\address{$^a$ Departamento de F{\'\i}sica Te\'orica y del Cosmos and CAFPE, \\
Universidad de Granada, E-18071 Granada, Spain \\
$^b$ Departament de Física Teòrica and IFIC, Universitat de València-CSIC, \\ 
Dr. Moliner 50, E-46100 Burjassot (València), Spain \\
$^c$ Department of Physics and Astronomy, University of California, Riverside \\ 
CA 92521-0413, USA} 
}
\maketitle
\begin{abstract}
We review the SM extensions with scalar multiplets including 
doubly-charged components eventually observable as 
di-leptonic resonances at the LHC. 
Special emphasis is payed to the limits on LNV implied by 
doubly-charged scalar searches at the LHC, 
and to the characterization of the multiplet doubly-charged 
scalars belong to if they are observed to decay into 
same-sign charged lepton pairs. 
\end{abstract}
\PACS{11.30.Fs, 12.60.Fr, 13.15.+g, 14.60.St, 14.80.-j 
}

% 11.30.Fs Global symmetries (e.g., baryon number, lepton number)
% 12.60.Fr Extensions of electroweak Higgs sector
% 13.15.+g Neutrino interactions
% 13.35.Hb Decays of heavy neutrinos
% 14.60.St Non-standard-model neutrinos, right-handed neutrinos, etc.
% 14.60.Pq Neutrino mass and mixing (see also 12.15.Ff Quark and lepton masses
%	 and mixing)
% 14.80.-j Other particles (including hypothetical)
  
\section{Introduction}

The Standard Model (SM) provides a quite precise description of 
particle physics up to the electro-weak (EW) scale. 
In particular, ATLAS \cite{Aad:2012tfa} and 
CMS \cite{Chatrchyan:2012ufa} have recently established 
the Brout-Englert-Higgs mechanism \cite{Englert:1964et,Higgs:1964pj}
providing to the SM fields a mass through the discovery of the 
Higgs boson, the last particle predicted by the SM. 
Moreover, not only no vestige of new physics (NP) has showed up 
at the LHC up to now, but EW precision data (EWPD) are 
in good agreement with the SM predictions to the 
radiative correction level \cite{delAguila:2011zs,deBlas:2013gla}. 
The SM also has two accidental global symmetries, baryon (B) and lepton (L) number: 
both anomalous but not their difference B$-$L. 
One may then wonder if they are also exactly realized in Nature, 
as predicted by the SM (up to non-perturbative effects). 
If broken, they are only very tinily violated. 
In fact, if the proton decays, B number would be broken, but 
we know that the proton mean life is extremely long 
$\tau_p > 2.1\times 10^{29}$ yr at 90\% C.L. \cite{Beringer:1900zz}. 
The observed 
B asymmetry of the universe is also quite small $\eta \sim 10^{-10}$ \cite{Beringer:1900zz}, 
as it is the B number violation required to explain it 
if this is actually its origin.  
Similarly, L number (LN) is only very tinily broken if it is not exact. 
The only low energy process which might provide conclusive 
evidence of LN violation (LNV), neutrinoless double beta decay 
($0\nu\beta\beta$), has not been undoubtedly observed 
$\tau_{\frac{1}{2}} ({^{76}{\rm Ge}} \rightarrow {^{76}{\rm Se}} + 2e^-) > 1.9 \times 10^{25}$ yr 
at 90\% C.L. \cite{Beringer:1900zz}. 
Besides, if the B asymmetry is due to leptogenesis \cite{Hambye:2012fh} and 
hence to LNV, its amount at low energy should be rather small, too. 

As a matter of fact, the only conclusive departure from the SM is 
neutrino oscillations \cite{Beringer:1900zz}, which is explained introducing neutrino 
masses in the model, but neutrinos may be Dirac, and LN 
conserved, or Majorana and in this case neutrino masses would provide 
the only conclusive evidence of LNV up to now. 
But this should be eventually tested through the consistency with other related 
experimental results, as, for instance, the observation of $0\nu\beta\beta$ 
with half-life $\gtrsim 10^{26}$ yr for $|m_{ee}| \sim$ few per cent eV \cite{Avignone:2007fu}. 
Although even if $0\nu\beta\beta$ is observed, the main contribution to this process could have 
a different source \cite{delAguila:2011gr,delAguila:2012nu}. 
However, at the LHC era the question is which is the LHC potential 
for observing LNV signals, as no B number violating (model independent) 
signature can emerge at the LHC.  

In the following, we want to argue first that although the observation of LNV  
at the LHC \cite{Keung:1983uu} may be problematic, it is quite plausible 
in definite models and parameter space regions. 
As a matter of fact, LHC is especially sensitive to doubly-charged 
scalars with two-body decays into leptons \cite{Chatrchyan:2012ya,ATLAS:2012hi}, 
which is the case we will concentrate on.  
Then, we review the expected limits on LNV in the first and in the next LHC run, 
and how to determine the type of multiplet the doubly-charged 
scalar belongs to.

In general, due to the smallness of LNV in low energy processes involving only SM particles, 
LNV effects must be banished to high energy or 
to a secluded sector, which may or not gauge LN 
\footnote{An interesting type of models gauging B and L is described in 
\cite{Duerr:2013dza}. They include two new $Z^\prime$s coupling 
to quarks and leptons, respectively, but their interactions do 
not violate LN in the SM sector. Right-handed (RH) neutrinos 
get Majorana masses, not necessarily heavy, through the vacuum 
expectation value (VEV) of a scalar singlet breaking LN; whereas 
LNV remains small in the SM processes.}. 
In the usual scenario with a new heavy sector with large 
LNV couplings, the effective operators obtained by integrating 
the heavy modes out and hence describing the NP but only involving 
SM particles are suppressed by the corresponding power of the large effective 
scale $\Lambda^n$. The lowest dimensional and formally 
dominant operator being the Weinberg operator 
$\mathcal{O}^{(5)} = (\overline{L^c_L}\tilde{\phi}^*)(\tilde{\phi}^\dagger L_L)$ 
\cite{Weinberg:1979sa} 
\footnote{$L_L = \left( \begin{array}{c} \nu_L \\ l_L \end{array} \right)$ and 
$\phi = \left( \begin{array}{c} \phi^+ \\ \phi^0 \end{array} \right)$ 
are the SM lepton and Higgs doublets, respectively.}. 
Which only gives Majorana masses to the 
SM neutrinos and couples the Higgs to neutrino pairs 
with |LN| = 2, although with very small LNV effective couplings.  

The simplest SM extensions generating this operator are 
the {\it see-saw} of type I \cite{seesawI}, II \cite{seesawII} and III \cite{seesawIII}. 
Type I and III are mediated by fermions transforming as an EW singlet and triplet, 
respectively, and type II by a scalar triplet. 
If these messengers have masses near the TeV, appropriate combinations of LNV couplings 
must be effectively small, conspiring not to provide the SM neutrinos too large a mass. 
As the production mechanism can not be suppressed if the extra 
heavy particles have to be produced copiously enough at the LHC, 
LN must be violated in their decays. This in turn implies that there must 
be at least two different channels, and none of them can dominate 
if LNV has to be observable.   
This restricts the model; what in general may appear to require fine tuning. 
But there are models where this seems not to be severe, 
and in any case we must assume this to be the case if we want to search for 
genuine LNV  signals. 

\begin{figure}[]
\begin{center}
\hspace{-0.5cm}
\begin{tabular}{cc}
\epsfig{file=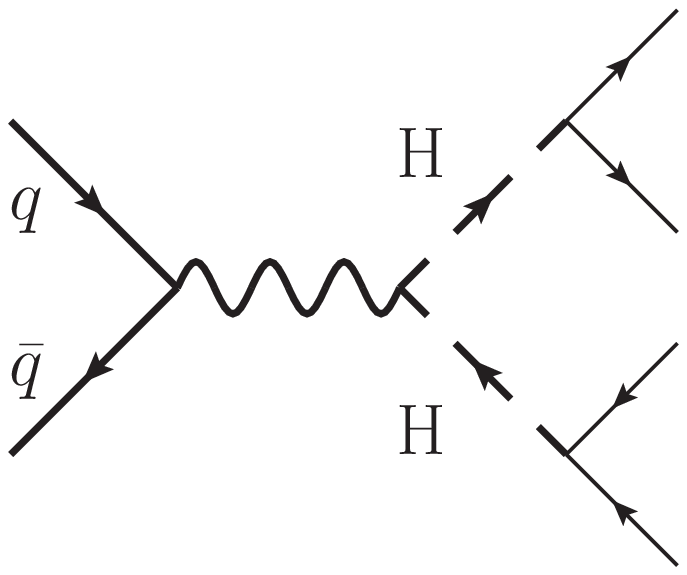,width=2.7cm,clip=} & 
\epsfig{file=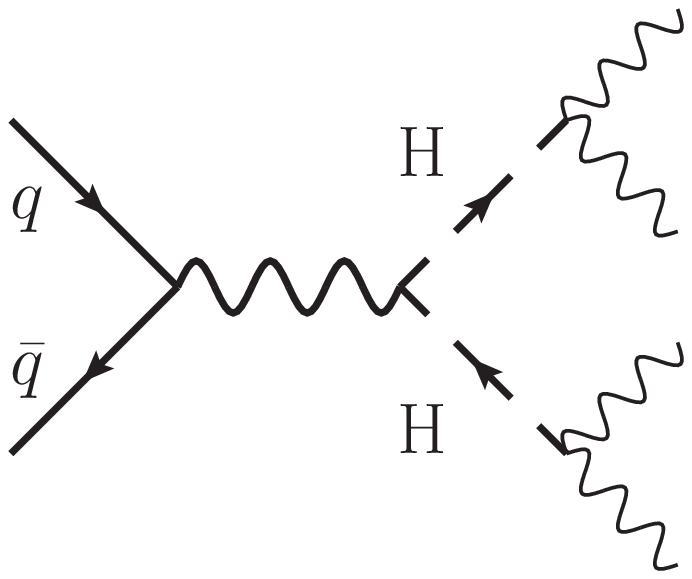,width=2.7cm,clip=} \\ [1mm] 
(a) & (b) \\ [1mm] 
{\small $l^+ l^+ l^- l^-, l^\pm l^\pm l^\mp\nu_l$} & 
{\small $W^+ W^+ W^- W^-, W^\pm W^\pm W^\mp Z$} \\ [1mm]
\epsfig{file=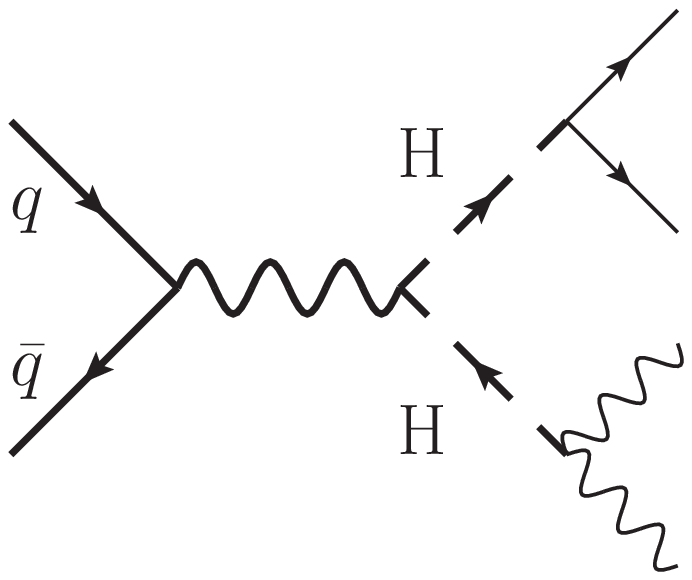,width=2.7cm,clip=} & 
\epsfig{file=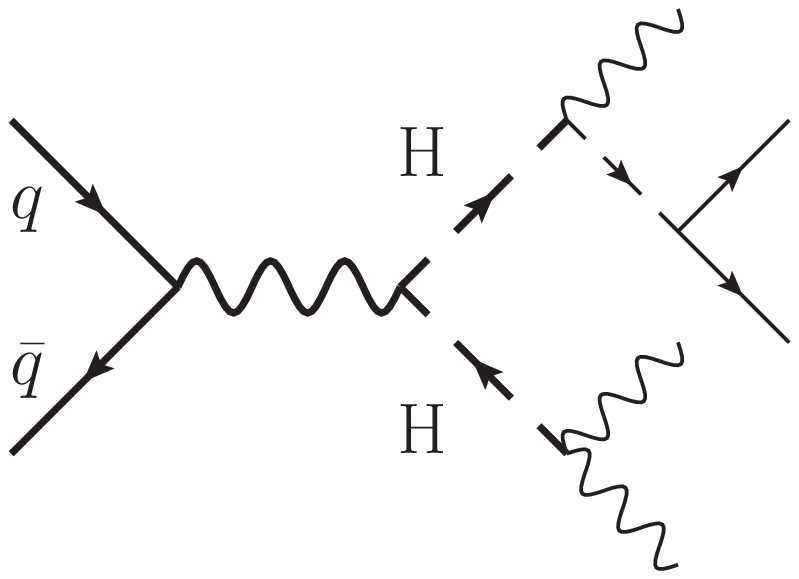,width=2.7cm,clip=} \\ [1mm]
(c) & (d)  \\ [1mm] 
{\small $l^\pm l^\pm W^\mp W^\mp, l^\pm l^\pm W^\mp Z$} & 
{\small $l^\pm l^\pm  W^\pm W^\mp W^\mp$} \\ [1mm]
\end{tabular}
\end{center}
\caption{(a), (b), (c): Doubly-charged scalar pair (${\rm H}^{++}{\rm H}^{--}$) and associated 
(${\rm H}^{\pm\pm}{\rm H}^{\mp}$) production diagrams. 
(d): Associated production with a triply-charged scalar (${\rm H}^{\pm\pm\pm}{\rm H}^{\mp\mp}$) 
when the multiplet also has a component with charge $|{\rm Q}| = 3$.}
\label{diagrams}
\end{figure}

\begin{figure}[]
\begin{center}
\begin{tabular}{cc}
\epsfig{file=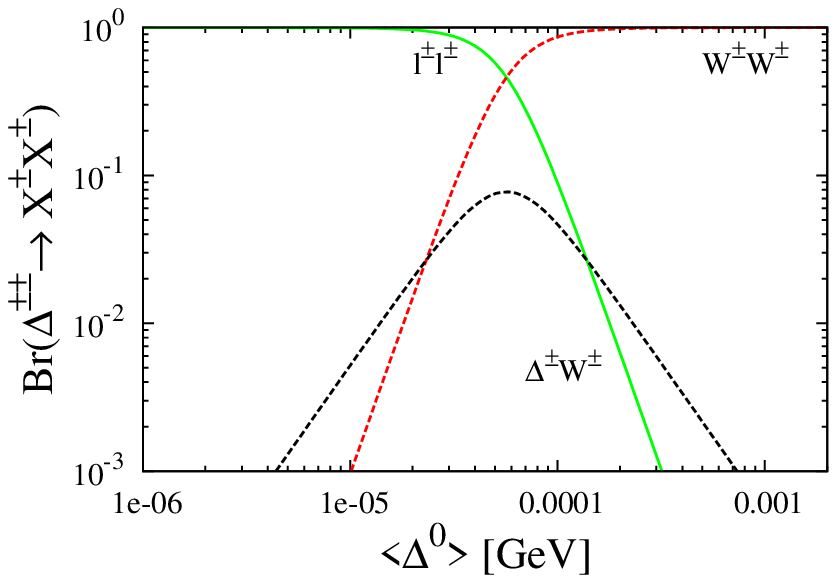,width=6.0cm,clip=} & 
\epsfig{file=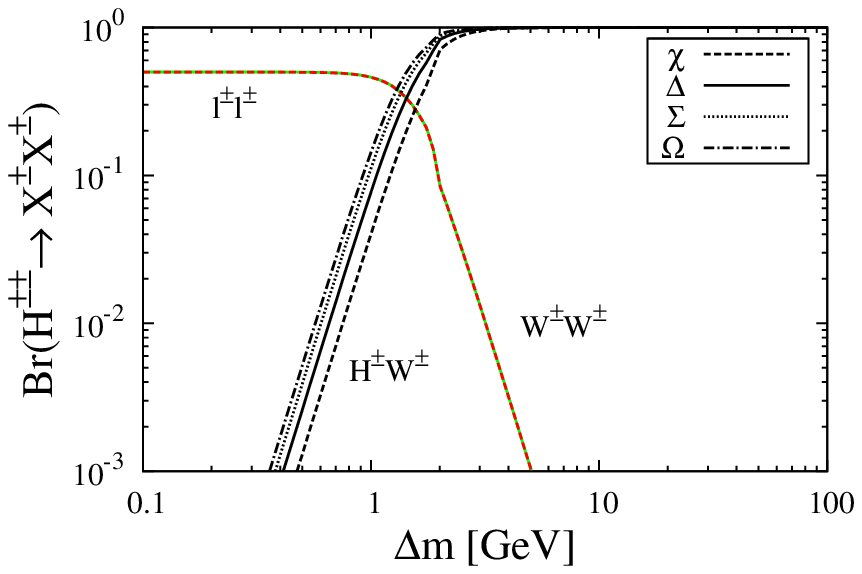,width=6.0cm,clip=} \\ 
\end{tabular}
\end{center}
\caption{Left: Scalar branching ratios for the triplet $\Delta$ as 
a function of $\langle \Delta^0 \rangle$ for 
$\sum_{i=1,2,3} m_{\nu_i}^2 = 0.1^2 \ {\rm eV}^2$ and $m_{\Delta^{\pm\pm}} = 500$ GeV, 
and $\Delta m = 1$ GeV for the decay into $\Delta^\pm W^{\pm *} \rightarrow \Delta^\pm e^\pm \nu_e, \cdots$. 
Right: Scalar branching ratios for different H multiplets as 
a function of $\Delta m = m_{{\rm H}^{\pm\pm}} - m_{{\rm H}^{\pm}}$ for $m_{{\rm H}^{\pm\pm}} = 500$ GeV and 
$\Gamma ({\rm H}^{\pm\pm} \rightarrow l^\pm l^\pm ) = \Gamma ({\rm H}^{\pm\pm} \rightarrow W^\pm W^\pm )$ 
and the ${\rm H} W_\mu W^\mu$ coupling equal to $5.5\times 10^{-5} g^2$ GeV.
}
\label{branchingratios}
\end{figure}

For instance, in the type II case the scalar triplet $\Delta$ is (pair and associated) 
produced with EW strength for it transforms non-trivially under the ${\rm SU(2)_L\times U(1)_Y}$ 
transformations; and can decay into lepton and boson pairs 
for it couples to two identical (neglecting family replication) lepton doublets 
(which defines its |LN| = 2) and to gauge boson pairs (with LN = 0) if its neutral 
component gets a VEV, $\langle \Delta^0 \rangle \neq 0$, breaking LN. 
Thus, if the Yukawa coupling is too large the triplet components always decay 
into two leptons (diagram (a) in Fig. \ref{diagrams}) \cite{Hektor:2007uu}; and if it is very small 
and $\langle \Delta^0 \rangle$ large enough, then their decay is always 
into two gauge bosons (diagram (b) in Fig. \ref{diagrams}). 
The LNV process with each of the two pair (associated) produced scalars 
decaying in a different mode (diagram (c) in Fig. \ref{diagrams}) is highly 
suppressed in both extreme cases. As in Ref. \cite{Perez:2008ha} we plot 
in Fig. \ref{branchingratios} (left) the two-body branching ratios for the 
doubly-charged scalar component as a function of the effective di-boson 
coupling (equal to $g^2\langle \Delta^0 \rangle$ if the scalar multiplet is a triplet) 
properly normalized (divided by $g^2$).  
As it is apparent, only near $\langle \Delta^0 \rangle \approx 5.5\times10^{-5}$ both 
decay rates are comparable and hence genuine LNV signals eventually observable 
at the LHC. 
This value is fixed in the {\it see-saw} of type II because the 
neutrino masses are proportional to both, the Yukawa couplings 
and to $\langle \Delta^0 \rangle$, and hence can be 
related to the messenger (doubly-charged scalar) decays into 
leptons, and into bosons if we require both branching ratios to be comparable 
\cite{Perez:2008ha,delAguila:2008cj}:
\begin{equation}
\begin{tabular}{l}
$\sum_{i,j=e,\mu,\tau} \Gamma (\Delta^{\pm\pm}\rightarrow l_i^\pm l_j^\pm) = 
\frac{m_\Delta}{8\pi} \frac{\sum_{i=1,2,3} m_{\nu_i}^2}{4\langle \Delta^0 \rangle^2}$ , \\
$\Gamma (\Delta^{\pm\pm}\rightarrow W^\pm W^\pm) = 
\frac{g^4 \langle \Delta^0 \rangle^2}{32\pi} \frac{m_\Delta^3}{m_W^4}$  $\quad {\rm in\ the\ limit} 
\ m_W^2 \ll m_\Delta^2$ .
\end{tabular}
\label{bound}
\end{equation}
Thus, fixed the scalar mass $m_{\Delta^{\pm\pm}} = 500$ GeV and the sum of the 
three SM neutrino masses $\sum_{i=1,2,3} m_{\nu_i}^2 = 0.1^2 \ {\rm eV}^2$, 
and assuming both decay rates equal, we can determine for which $\langle \Delta^0 \rangle$ 
both curves cross in Fig. \ref{branchingratios} (left). 

These relations need not apply in more elaborated SM extensions but 
realistic models must accommodate the correct SM neutrino masses which in general 
constrains the size of LNV in the SM sector. 
In the {\it see-saw} of type II it is enough to satisfy the limits on neutrino 
masses to fulfill the bounds on $0\nu\beta\beta$, 
but this is not the case in general \cite{delAguila:2011gr,delAguila:2012nu}. 
The corresponding constraint implied by the limit on $0\nu\beta\beta$, 
as well as the EW limit on $\langle \Delta^0 \rangle$ not to upset the 
bounds on the $\rho$ parameter, or the requirement of producing enough 
leptogenesis if this is the origin of the observed B number asymmetry of the universe, 
must be also satisfied by realistic SM extensions. 

In order to observe LNV at the LHC not only the two types of couplings 
involved in the process must be of the same order but no other messenger 
decay can be much larger. What further constrains the model, 
restricting the mixing of the new heavy multiplets with other scalars to small values. 
Taking again the {\it see-saw} of type II as example, we plot in 
Fig. \ref{branchingratios} (right) the $\Delta^{\pm\pm}$ branching 
ratios into $l^\pm l^\pm$ and $W^\pm W^\pm$ assuming that 
they are equal and into $\Delta^\pm W^{\pm *}$ as a function of 
the mass splitting between contiguous components 
$\Delta m = m_{\Delta^{\pm\pm}} - m_{\Delta^{\pm}}$, 
which completely dominates for differences larger than 
the GeV \cite{Perez:2008ha,Grifols:1989xe} 
\footnote{$\Gamma ({\rm H}^{\pm\pm}\rightarrow {\rm H}^\pm W^{\pm *}
\rightarrow {\rm H}^\pm e^{\pm}\nu_e , \cdots) = 
n\frac{g^4 {\rm T}}{240\pi^3} \frac{\Delta m^5}{m_W^4}$ for small $\Delta m$, 
where $n$ is the number of open lepton and quark channels 
(5 for $\Delta m < m_\tau, \cdots$) and T the total isospin 
(0 for $\kappa$, $\frac{1}{2}$ for $\chi$, 1 for $\Delta$, $\cdots$).}. 
 On the left we assume $\Delta m = 1$ GeV 
(and $m_{\Delta^{\pm\pm}} = 500$ GeV). Then, in general, 
the mixing between different scalar multiplets must be rather small.   

These considerations have different consequences when the mediator is 
a heavy Majorana neutrino transforming as a singlet or as a member of a triplet. 
In both cases the two different decay modes are 
conjugated from each other, and hence of equal size 
\footnote{Although the corresponding decay rates are equal at leading 
order (tree level), they can differ at higher order (one loop) and give rise to 
the corresponding CP asymmetry, eventually generating enough 
leptogenesis \cite{Hambye:2012fh}. At any rate, 
NP signals are often simulated using tree-level amplitudes, 
as in our case, thus neglecting possible CP violating effects.}. 
The required suppression for a TeV messenger comes in this case 
from small mixing angles 
or large interferences as in the quasi-Dirac case \cite{delAguila:2008hw}. 
As singlets must be produced through mixing, 
the LHC reach is very much reduced \cite{Han:2006ip} 
\footnote{Even if vector-boson fusion contributions are large \cite{Dev:2013wba}, 
present limits on lepton mixing make difficult to observe this process 
\cite{deBlas:2013gla,delAguila:2008pw}.}. 
In any case, the picture can be modified as soon as a new layer of 
NP is reached. 
For example, if parity is restored at higher energy \cite{LR} 
and RH charged gauge bosons with several TeV masses 
are produced, their LNV decays can be unsuppressed and 
observable \cite{WR_N}. The phenomenological 
constraints on neutrino oscillations, $0\nu\beta\beta$ and 
leptogenesis are still present but the suppression factors 
combine several effects which can enter differently in the 
LHC production process. 

Having discussed the difficulties for observing LNV directly at the LHC, 
let us follow the optimistic approach and consider how large 
the LHC reach can be. If the new sector does not have strong 
interactions and mixes little, the LNV mediator has to be pair produced 
and hopefully with EW strength. Moreover, in order to 
ease its reconstruction, it must resonate in two-body channels 
because then final products will have larger momenta, much less probable 
within the SM. These conditions uniquely characterize 
doubly-charged scalars, but not the full multiplet H they belong to. 
They can be part of a triplet, as in the {\it see-saw} of type II, 
or of an EW multiplet with arbitrary isospin 
\cite{delAguila:2013yaa,delAguila:2013mia}. In the following we compare 
the LHC potential for the different EW multiplet quantum number assignments. 
For a general multiplet, the larger the total isospin T and hypercharge Y, the larger is 
the charge Q = T$_3$ + Y that the components can have. We will 
restrict ourselves to multiplets for which the highest charge is 2 
($|{\rm Q}_{\rm max}|=2$): ${\rm T}_{\rm Y} =$ {\bf 0}$_2$ (singlet $\kappa$), 
$\mathbf{\frac{1}{2}}_{\frac{3}{2}}$ (doublet $\chi$), {\bf 1}$_1$ (triplet $\Delta$), 
$\mathbf{\frac{3}{2}}_{\frac{1}{2}}$ (quadruplet $\Sigma$), 
{\bf 2}$_0$ (quintuplet $\Omega$). 
Multiplets with higher charges \cite{Babu:2009aq} can have striking 
signatures, 
${\rm H}^{\pm\pm\pm} \rightarrow {\rm H}^{\pm\pm} W^\pm \rightarrow {\ell}^{\pm}{\ell}^{\pm} W^\pm $ 
(see, for example, Fig. \ref{diagrams} (d)) but the momenta of the final products 
are smaller and one must identify doubly-charged resonances in any case, 
although the total cross-section is in general also larger for larger T 
(it also depends on T$_3$ and Y \cite{delAguila:2013mia}). 
Not only the production cross-sections but the decays within the multiplet 
depend on the component quantum numbers. 
In Fig. \ref{branchingratios} we also plot the 
${\rm H}^{\pm\pm} \rightarrow {\rm H}^{\pm} W^{\pm *}$ branching ratio, 
which grows with T, for the different multiplets the doubly-charged scalar can belong to above.   

\section{LHC bounds on doubly-charged scalar masses \break\hfill 
and their LNV signals}

CMS \cite{Chatrchyan:2012ya} and ATLAS \cite{ATLAS:2012hi} 
have searched for doubly-charged scalars decaying into electrons 
and muons, setting stringent bounds on their mass. These, 
however, are very much dependent on the doubly-charged scalar 
branching ratios, as it is apparent from the CMS analyses allowing 
for $\Delta^{\pm\pm}$ decays into tau leptons. 
Obviously, this is even more dramatic if the gauge boson channel 
$W^\pm W^\pm$ is also sizable, which is compulsory in order to observe 
LNV, as emphasized above. 
To perform searches for LNV signals mediated 
by doubly-charged scalars at the LHC, 
$pp \rightarrow {\rm H}^{\pm\pm} {\rm H}^{\mp\mp}, {\rm H}^{\pm\pm} {\rm H}^{\mp} 
\rightarrow {\ell}^{\pm}{\ell}^{\pm} W^\mp W^\mp ,  {\ell}^{\pm}{\ell}^{\pm} W^\mp Z$, 
we have implemented the corresponding models in \texttt{MadGraph5} \cite{Alwall:2011uj}, 
which can be obtained under request \cite{Request}. 
(A full description of the Monte Carlo implementation and the analysis 
described in this section can be found in Ref. \cite{delAguila:2013mia}.)
Using it we can mimic the CMS analyses and estimate 
the corresponding bounds. 
In Table \ref{table:Bounds} we collect in the first row, for 
$\sqrt s$ = 7 TeV and ${\mathcal{L}_{\rm int}} = 4.9\ {\rm fb}^{-1}$, 
the corresponding limit on ${\rm m}_{{\Delta}^{\pm\pm}} \sim 400$ GeV, 
assuming that doubly-charged scalars 
decay 100 \% of the time into $\ell^{\pm}\ell^{\pm}$, $\ell = e, \mu$.
\begin{table}[]
\begin{center}
{ 
\begin{tabular}{ c || c c c c c } 
& \multicolumn{5}{c}{${\mathbf{Isospin}_{\rm hypercharge}}$} \\
\vspace{-0.3cm}
&&&&&\\
$\sqrt s$ (TeV), ${\mathcal{L}_{\rm int} ({\rm fb}^{-1})}$ 
& \quad ${\mathbf{0}_{2}}$  \quad & \quad ${\mathbf{\frac{1}{2}}_{\frac{3}{2}}}$  \quad 
& \quad ${\mathbf{1}_{1}}$  \quad & \quad ${\mathbf{\frac{1}{2}}_{\frac{3}{2}}}$ \quad 
& \quad ${\mathbf{2}_{0}}$ \quad \\
\vspace{-0.35cm}
&&&&&\\
\hline
\hline
\vspace{-0.35cm}
&&&&&\\
& \multicolumn{5}{c}{${\rm m}_{{\rm H}^{\pm\pm}}$ (GeV) bounds from $\ell^\pm \ell^\pm \ell^\mp \ell^\mp$} \\ 
\vspace{-0.35cm}
&&&&&\\
7\;,\; 4.9 & 340 & 350 & 395 & 450 & 490 \\
8\;,\; 20 & 480 & 490 & 550 & 610 & 665 \\
14\;,\; 100 & 900 & 915 & 1030 & 1140 & 1230 \\
\hline
\vspace{-0.35cm}
&&&&&\\
& \multicolumn{5}{c}{${\rm m}_{{\rm H}^{\pm\pm}}$ (GeV) bounds from $\ell^\pm \ell^\pm W^\mp W^\mp$} \\ 
\vspace{-0.35cm}
&&&&&\\ 
7\;,\; 4.9 & < 200 & < 200 & < 200 & 230 & 335 \\
8\;,\; 20 & 240 & 260 & 340 & 415 & 475 \\
14\;,\; 100 & 540 & 570 & 720 & 850 & 940 \\
\hline
\vspace{-0.35cm}
&&&&&\\
& \multicolumn{5}{c}{${\rm m}_{{\rm H}^{\pm\pm}}$ (GeV) bounds from $\ell^\pm \ell^\pm W^\mp Z$} \\ 
\vspace{-0.35cm}
&&&&&\\
7\;,\; 4.9 & $-$ &  < 200 &  < 200 &  < 200 &  < 200 \\
8\;,\; 20 & $-$ & 210 & 280 & 330 & 360 \\
14\;,\; 100 & $-$ & 470 & 620 & 720 & 780 \\
\end{tabular}
\caption{\label{table:Bounds}
Estimated limits on the cross-section and on the corresponding 
scalar mass ${\rm m}_{{\rm H}^{\pm\pm}}$ (GeV) as a function of 
the multiplet if belongs to from LHC searches for doubly-charged 
scalars. 
The $\ell^\pm \ell^\pm \ell^\mp \ell^\mp$ analysis 
assumes that 
${{\rm H}^{\pm\pm}} \rightarrow \ell^\pm \ell^\pm$ 
100 \% of the time; whereas the other two analyses assume 
a 50 \% branching ratio for each decay mode 
of ${{\rm H}^{\pm\pm}} \rightarrow \ell^\pm \ell^\pm , W^\pm W^\pm$ 
and 
of ${{\rm H}^{\pm}} \rightarrow \ell^\pm \nu_\ell , W^\pm Z$. 
}}
\end{center}
\end{table}
Obviously, this bound depends on the (pair) production cross-section 
which in turn depends on the EW multiplet the doubly-charged 
scalar belongs to. In the same row we quote the corresponding 
bounds for the scalar multiplets H with no components of higher charge. 
As the cross-section grows with T so does the limit. 
In the other two rows below we estimate the bounds for 
$\sqrt s$ = 8 and 14 TeV and ${\mathcal{L}_{\rm int}} = 20$ and 
100 ${\rm fb}^{-1}$, respectively \cite{delAguila:2013mia}. For the three energies we apply the 
same cuts as CMS for 7 TeV. Certainly these cuts will be optimized 
by the LHC collaborations for higher energies and hence the corresponding 
bounds improved, but they should not differ much from our estimates 
if no event excess is observed. 

This analysis can be extended to scalar decays into gauge bosons and 
hence to LNV signals.  
We collect the corresponding bounds on 
${\ell}^{\pm}{\ell}^{\pm} W^\mp W^\mp$ and ${\ell}^{\pm}{\ell}^{\pm} W^\mp Z$ 
production in the second and third sets of rows of the Table, respectively, 
for the different LHC runs and doubly-charged scalar multiplet assignments \cite{delAguila:2013mia}. 
We assume for these analyses the same set of cuts applied by CMS for 
$\Delta$ decaying into $\ell\tau$ at 7 TeV, and that the number of 
observed events coincides with the background estimate. 
In both channels we also assume that the heavy scalars decay 50 \% 
of the time in each of the two modes (electron or muon and gauge boson pairs), 
being all other possible decay modes negligible, especially the cascade 
decays within the scalar multiplet. 
These analyses are based on four and three-lepton samples \cite{Chatrchyan:2012ya}. 
Although the estimates for LNV only use the three-lepton sample, 
which is much more sensitive than the four-lepton one to final modes involving also gauge 
bosons for pair and associated scalar production, 
with efficiencies almost an order of magnitude larger. 
In the last three rows there is no bound for the singlet because 
it does not have a singly-charged component and hence the doubly-charged 
scalar can be only pair produced. As no experimental analysis 
is available for scalar masses below 200 GeV, we write < 200 GeV 
in the Table when there is no enough sensitivity to set a bound 
for larger masses. 

\section{Doubly-charged scalar multiplet determination at the LHC}

Doubly-charged scalars are predicted by many SM extensions 
and may show up at the LHC even if no LNV signal can be ever 
established at colliders. Therefore, a resonance in the same-sign 
charged di-lepton channel can be detected and hence the question 
is if the EW multiplet it belongs to can be determined. 
As explained above, the production cross-section depends 
on the total isospin and hypercharge but the number of observed events  
in each final state is also proportional to the corresponding branching ratio. 
Then, a multi-sample analysis is mandatory. In \cite{delAguila:2013yaa} 
we have proposed how to measure the doubly-charged scalar pair production 
cross-section and hence how to determine the multiplet it belongs to, assuming 
that only two-body decays are sizable and the two-lepton channel 
${\rm H}^{\pm\pm} \rightarrow {\ell}^{\pm}{\ell}^{\pm}$ is observable. 
However, only with a relatively large statistics and a large enough 
${\rm H}^{\pm\pm} \rightarrow {\ell}^{\pm}{\ell}^{\pm}$ branching ratio 
it is possible to obtain a crucial test 
(see Ref. \cite{delAguila:2013yaa} for a detailed quantitative discussion). 
For example, the production cross-sections for the different multiplets stay apart 
by at least 3 $\sigma$ if ${\rm H}^{\pm\pm}$ only decays into ${\ell}^{\pm}{\ell}^{\pm}$ 
for $\sqrt s =$ 14 TeV and ${\mathcal{L}_{\rm int}} = 300\ {\rm fb}^{-1}$. 
However, if ${\rm H}^{\pm\pm}$ decays 50 \% of the time into 
${\ell}^{\pm}{\ell}^{\pm}$ and ${\ell}^{\pm}{\tau}^{\pm}$, respectively, 
this integrated luminosity is not enough to separate the doublet from the triplet, 
and a longer run to accumulate $3000\ {\rm fb}^{-1}$ becomes 
necessary to distinguish the different cases.

\vspace{0.5cm}
We thank A. Aparici and J. Santiago for useful discussions and the
careful reading of the manuscript. 
This work has been supported in part by the Ministry of Economy and 
Competitiveness (MINECO), under the grant numbers FPA2006-05294, 
FPA2010-17915 and FPA2011-23897, by the Junta de Andaluc{\'\i}a 
grants FQM 101 and FQM 6552, by the 
``Generalitat Valenciana'' grant PROMETEO/2009/128, and by the 
U.S. Department of Energy grant No.~DE-FG03-94ER40837. 
M.C. is supported by the MINECO under the FPU program.

\end{document}